\documentstyle[pre,aps,epsf,amssymb]{revtex}

\newcommand{\figurewidth}{8.5cm}

\begin{document}

\draft % makes pacs numbers print

\title{Crossover behavior in $^3$He and Xe near their liquid--vapor
       critical point}
\author{Erik Luijten$^{1,2,}$\thanks{Present address: Institute for Physical
        Science and Technology, University of Maryland, College Park, MD
        20742-2431, USA} and
        Horst Meyer$^{3}$}
\address{$^{1}$Max-Planck-Institut f\"ur Polymerforschung, Postfach 3148,
         D-55021 Mainz, Germany}
\address{$^{2}$Institut f\"ur Physik, WA 331, Johannes Gutenberg-Universit\"at,
         D-55099 Mainz, Germany}
\address{$^{3}$Department of Physics, Duke University, Durham, NC 27708-0305,
         USA}

\date{January 25, 2000}

\maketitle

\begin{abstract}
  We present a detailed discussion of the crossover from mean-field to Ising
  critical behavior upon approach of the critical point, both for $^3$He and
  Xe. By combining different sets of experimental data, we are able to cover an
  unusually large temperature range on either side of the critical
  temperature~$T_c$. Below $T_c$, we thus can make an accurate comparison with
  a recent calculation for the crossover of the coexistence curve. For the
  regime above $T_c$, an analysis of the compressibility demonstrates that the
  crossover regime in $^3$He is unexpectedly widened by a subtle interplay
  between quantum and critical fluctuations.
\end{abstract}


Recently, there has been a renewed interest in the nature of the crossover from
mean-field-like (``classical'') to Ising-like critical behavior that occurs
upon approach of the critical point. Although this phenomenon can be observed
in a wide variety of experimental systems, including simple fluids, micellar
solutions, and polymer mixtures, much of the recent attention has focused on
its theoretical description. An important reason for this lack of experimental
data is the width of the crossover region, which extends over several decades
in the reduced temperature $t \equiv (T-T_c)/T_c$, where $T_c$ is the critical
temperature. The crossover depends on the ratio between $t$ and the
(system-dependent) Ginzburg number~$G$: Ising critical behavior occurs for $t
\ll G$ and mean-field critical behavior is expected for $t \gg G$. Throughout
the crossover region, however, one has to stay within the critical regime,
i.e., $t \lesssim 0.1$. Hence, the full crossover can be observed only if $G$
is sufficiently small.  On the other hand, if $G$ is {\em extremely\/} small,
as for conventional superconductors which have $G \approx 10^{-16}$, the
nonclassical region is so narrow that it becomes impossible to observe.
Ideally, thus, one would need a system with a tunable Ginzburg number, as can
actually be realized in polymer mixtures, where the Ginzburg number is
inversely proportional to the molecular weight.  Measurements of the crossover
have indeed been reported for such
systems~\cite{schwahn87-bates90,meier93-schwahn94}, essentially confirming the
existence of a crossover between the two universality classes.  However, few
studies have actually addressed the {\em shape\/} of the crossover curves.  In
Ref.~\cite{meier93-schwahn94} the concentration susceptibility $\chi$ as a
function of $t/G$ was shown to be well described by a phenomenological
crossover function obtained by Belyakov and Kiselev (BK)~\cite{belyakov92}, but
it must be pointed out that $\chi$ increases by more than five orders of
magnitude in the crossover region, making it difficult to judge even the
qualitative agreement from a logarithmic plot.  The {\em effective\/}
susceptibility exponent, which is defined as the logarithmic derivative of
$\chi$, $\gamma_{\rm eff} \equiv - d\ln \chi / d \ln |t|$~\cite{kouvel64}, is
clearly a much more sensitive quantity. However, very few
papers~\cite{bagnuls84b,anisimov95} have endeavored to discuss the shape of the
crossover in experimental systems in terms of this parameter.

It is the objective of this work to present an alternative description of
experimental data exhibiting (part of) the crossover between mean-field-like
and Ising-like (asymptotic) critical behavior. Our description differs from
previous ones in several aspects. Firstly, those of
Refs.~\cite{belyakov92,anisimov95} possess essentially phenomenological
features and rely on extensions of low-order renormalization-group~(RG)
results~\cite{belyakov92} or on the so-called RG matching
procedure~\cite{anisimov95}. Note that for simple liquids (as studied in this
work) the crossover function used in~\cite{anisimov95} closely resembles the
equation obtained by BK (cf.\ Ref.~\cite{anisimov92}).  In addition, both the
field-theoretic results obtained in Ref.~\cite{bagnuls84a-85} and the
description by BK are only valid in the limit where $t/G$ is varied but where
both $t \to 0$ and $G \to 0$, a restriction that is certainly not fulfilled by
liquids and liquid mixtures, for which $G$ is a finite, fixed quantity.  In
contrast, we use here theoretical crossover curves that have been obtained from
numerical calculations for model systems in which $G$ was a tunable
parameter~\cite{chi3d}. A second aspect concerns our choice of experimental
systems, {\em viz.\/}\ $^{3}$He and~Xe.  Whereas the latter system was already
considered for $t>0$ in Refs.~\cite{bagnuls84b,anisimov95}, it is the former
that exhibits two features that make it stand apart from other fluids: (i) its
coexistence curve is almost symmetrical with respect to $\rho_c$ in the
$\rho$--$|t|$ plane, thereby eliminating important corrections to scaling; (ii)
the magnitude of its bare correlation length suggests a relatively small
Ginzburg number.  In addition, there is a great abundance of available data
close to the liquid--vapor critical point for these two systems, extending over
a larger range in $t \gtrless 0$ than for most other fluids. They can be
thought of as representing extremes in both mass density and molar mass for
simple fluids near the critical point, $^3$He having the largest de Boer
parameter $\Lambda^*$ of any fluid~\cite{Hirschfelder}, implying a
``near-quantum'' behavior, and Xe a very small one, leading one to expect
``classical'' (in the sense of non-quantum) behavior. Both, however, show the
predicted critical exponents and amplitude ratios in the asymptotic critical
regime.  Our comparison now allows the exploration of the influence of quantum
effects on the nature of the crossover.  A final noteworthy aspect is the fact
that we consider the two-phase region $T<T_c$ as well. To our knowledge, this
is one of the first comparisons between a theoretical description of the
crossover of the coexistence curve and experimental results presented together
with the crossover above $T_c$. An important feature of the low-temperature
region is that one can clearly identify the end of the critical regime. By
contrast, for $T>T_c$, the region where the compressibility exhibits
mean-field-like {\em critical\/} behavior, i.e., $\gamma_{\rm eff} = 1$, cannot
be distinguished from the regular high-temperature behavior $\chi \sim 1/T$.
Thus, a system that actually has left the critical region can incorrectly be
interpreted as having completed the full crossover. An important point in this
context is the expected degree of universality of crossover scaling functions,
which may be legitimately doubted because of the decrease of the correlation
length in the crossover regime.  See Ref.~\cite{pelissetto99} for a detailed
analysis of the nature of nonuniversal corrections. Here we just note that
recent numerical work~\cite{dblock} has provided evidence that the actual
crossover scaling functions may be remarkably insensitive to the detailed
nature of the interactions.

For the study of crossover phenomena in $^3$He and Xe, we consider two
properties, namely the isothermal compressibility $\kappa_T$=
$\rho^{-1}(\partial\rho/\partial P)_T$ along the critical isochore above $T_c$
and the coexistence curve below $T_c$.  Here $\rho$ is the mass density and $P$
the pressure. The data are presented in the form of the dimensionless
``reduced'' compressibility $\chi^*_T$ = $P_c \kappa_T$ and (for Xe) the
coexistence-curve diameter $\Delta\rho^* = (\rho_{\rm liq} -\rho_{\rm
vap})/2\rho_c$ as a function of~$|t|$.  The densities are those of the
coexisting liquid and vapor phases.  This representation eliminates the effect
of the slope of the rectilinear diameter.  For $^3$He, where this slope is much
smaller than in Xe (cf.\ Fig.~7 of Ref.~\cite{Pittman:D:M:79}), we use the
individual liquid and vapor data in the form $\Delta\rho_+^* = (\rho_{\rm
liq}-\rho_c)/\rho_c$ and $\Delta\rho_-^* = (\rho_c - \rho_{\rm vap})/\rho_c$.
Inevitably, some accuracy is lost if one extracts effective exponents by means
of numerical differentiation of the actual data, and hence we have divided out
the leading singularity instead, which yields an equally sensitive comparison
with the theoretical models. Thus, the quantities are plotted as $\chi^*_T/t^{-
\gamma}$ and $\Delta\rho^*/(-t)^{\beta}$ with $\gamma$ = 1.240 and $\beta$ =
0.327, the asymptotic singular exponents predicted for the 3D Ising
model~\cite{Guida:ZJ:98}.  In a fully logarithmic plot versus $|t|$, the local
slopes are then, respectively, ($\gamma -\gamma_{\rm eff}$) and ($\beta_{\rm
eff} -\beta$). The influence of the uncertainties in the exponents $\beta$ and
$\gamma$ and in the respective critical temperatures of the examined fluids
turns out to be small.

For Xe, the critical parameters are $P_c= 58.40$ bar, $\rho_c = 1.110$
g/cm$^3$, $T_c= 289.73$ K. Excellent compressibility data along the critical
isochore have been obtained from light scattering experiments by G\"uttinger
and Cannell~\cite{Guettinger:C:81} for $10^{-4}< t < 10^{-1}$. The range of
these data overlaps with that covered by density measurements versus pressure
along several isotherms between $25\,^\circ{\rm C}$ and $150\,^\circ{\rm C}$ by
Michels {\it et al.}~\cite{Michels:W:L:54}.  From the tabulated data,
$\chi_T^*$ was obtained by numerical differentiation. A small (systematic)
adjustment of $4.5\%$ was required to bring the results in line with those of
Ref.~\cite{Guettinger:C:81}.  The combined range for Xe extends over
$3\frac{1}{2}$ decades in $t$ and exhibits a clear crossover in $\gamma_{\rm
eff}$ as we shall see.  For $\Delta \rho^* (-t)$, we used the data of
Ref.~\cite{Cornfeld:C:72}, where the densities of both the coexisting phases
were obtained simultaneously with a visual method. These data cover the range
$3.0\times 10^{-3} <-t< 2.9\times 10^{-1}$. Closer to the critical point, we
used the most recent data~\cite{Naerger:B:90}, obtained by an optical
interferometric technique.  In addition to a high precision, these data exhibit
a good agreement with those of Hocken and Moldover~\cite{Hocken:M:76} in the
asymptotic regime ($|t| < 5\times 10^{-5}$) and are consistent with those from
NMR experiments by Hayes and Carr~\cite{Hayes:C:77}.

For $^3$He ($P_c = 1.168$ bar, $\rho_c = 0.0414$ g/cm$^3$, $T_c = 3.317$ K on
the $T_{90}$ scale), compressibility data above $T_c$ were obtained from
density measurements along isotherms over the range $0.05 < t \lesssim
2$~\cite{Agosta:W:C:M:87} and from measurements of the vertical density
gradient in the gravitational field~\cite{Pittman:D:M:79}. For the latter
experiments, the determination of $\chi_T$ is limited to the temperature range
where the density gradient over the distance~$h$ is constant within the
experimental error.  As the stratification diminishes with increasing~$T$, the
scatter increases and the useful data were confined to $4\times 10^{-4} < t <
3\times 10^{-2}$. The data are consistent with earlier
ones~\cite{Wallace:M:70,Chase:Z:76}.  In both experiments, the
dielectric-constant method was used.  Also for the coexistence curve
$\Delta\rho^*(-t)$ the results of Ref.~\cite{Pittman:D:M:79} were used, where
the density of both phases could be measured simultaneously.  The range of the
data was $3\times 10^{-5}< -t \lesssim 0.1$ and could be extended by
measurements of $\rho_{\rm liq}$ at saturated vapor below 3.2 K ($0.05 < -t <
1.0$)~\cite{Kerr:T:62}, and of the vapor density $\rho_{\rm vap}$ in
equilibrium with the liquid above 1.4 K~\cite{Kerr:54}.  These measurements
used a volumetric method and join on smoothly to those of
Ref.~\cite{Pittman:D:M:79}.  Below 1.4 K, $\rho_{\rm vap}$ was obtained by
using the ideal gas law as a reasonable approximation.
 
\begin{figure}
\centering\leavevmode
\epsfxsize \figurewidth
\epsfbox{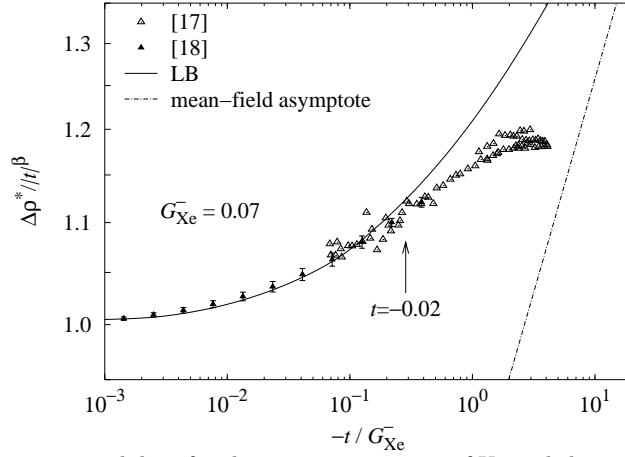}
\caption{Comparison between experimental data for the coexistence curve of Xe
  and the predicted universal crossover function calculated by Luijten and
  Binder~(LB). The vertical arrow calibrates the scale of~$t$. The data were
  normalized by dividing out the critical amplitude $B_{\rm Xe}=1.475$.}
\label{fig:xe-coex}
\end{figure}

We now proceed with a comparison of these experimental data to theoretical
predictions, starting with the coexistence curves. Figure~\ref{fig:xe-coex}
displays $\Delta\rho^* / |t|^\beta$ for Xe. The solid curve indicates the
crossover function as predicted from numerical data for the Ising-like systems
studied in Ref.~\cite{chi3d}. The closed symbols, which were recreated from the
fit expression of Ref.~\cite{Naerger:B:90} with parameters that were taken as
the average for the two samples, follow the predicted curve very closely.
These data are joined smoothly by those of Ref.~\cite{Cornfeld:C:72}, making
the data span a total range of approximately 80 K. For $|t| \lesssim 0.02$
there is good agreement with the theoretical curve; for $|t| > 0.02$ the latter
continues its trend toward the mean-field asymptote whereas the experimental
data already start to leave the critical region, presumably because of the
finite width of the low-temperature regime.  Indeed, the distance between the
onset of the deviation and absolute zero is less than two decades on this
logarithmic scale.  Note that there is only one adjustable parameter in the
form of the Ginzburg parameter $G^{-}_{\rm Xe}$, estimated as $0.07 \pm 0.02$.

\begin{figure}
\centering\leavevmode
\epsfxsize \figurewidth
\epsfbox{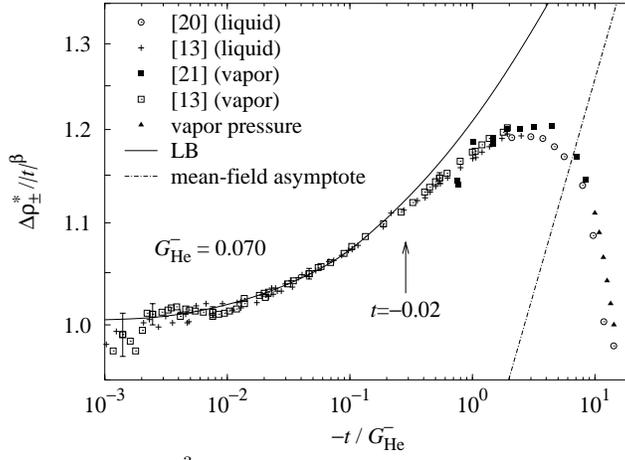}
\caption{Crossover of the coexistence curve for $^3$He. We adopted the critical
amplitude $B_{\rm He}=1.000$ and hence no further normalization of the data was
required.}
\label{fig:he-coex}
\end{figure}

The data for $^3$He, shown in Fig.~\ref{fig:he-coex}, exhibit a close
similarity to those for Xe. For $10^{-4} < |t| < 0.02$ the agreement with the
prediction is comparable with that for Xe and for larger $|t|$ a similar
changeover to regular low-temperature behavior is observed. The downward trend
upon approach of absolute zero simply results from the saturation of the order
parameter.  The Ginzburg parameter turns out to be the same as that for Xe
within the fitting uncertainty, with $G^{-}_{\rm He} = 0.070 \pm 0.008$.  The
close agreement of the experimental data for both Xe and $^3$He with the
theoretical curve confirms the expectation that, due to the predominance of the
critical fluctuations, the initial crossover behavior is essentially unaffected
by quantum statistics.  Unfortunately, the very small width of the critical
regime makes it impossible to identify the influence of quantum effects on the
crossover behavior at somewhat larger distances from the critical point.

\begin{figure}
\centering\leavevmode
\epsfxsize \figurewidth
\epsfbox{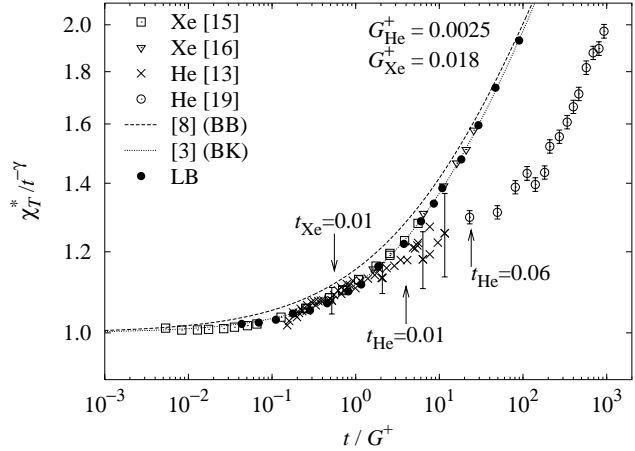}
\caption{Crossover of the compressibility of Xe and $^3$He for $T>T_c$. For Xe,
  the error bars do not exceed the symbol size. The data were
  normalized by dividing out the respective critical amplitudes $\Gamma^+_{\rm
  Xe}=0.0594$ and $\Gamma^+_{\rm He}=0.139$.}
\label{fig:compr}
\end{figure}

Next, we turn our attention to the high-temperature regime, where the crossover
of the compressibility upon approach of $T_c$ is considered. We compare the
experimental data to three theoretical descriptions, namely the
field-theoretical approach of Bagnuls and Bervillier (BB)~\cite{bagnuls84a-85},
the phenomenological extension of low-order RG calculations of
Ref.~\cite{belyakov92} and the numerical calculation of Luij\-ten and Binder
(LB)~\cite{chi3d}. The latter calculation has been improved compared to
Ref.~\cite{chi3d} by leaving out the results for systems with very short
interaction ranges, which cannot exhibit the crossover to mean-field-like
behavior due to their location with respect to the Wilson--Fisher RG fixed
point.  The improved calculation exhibits a very close agreement with the BK
curve.  In view of the phenomenological nature of the latter, this must be
coincidental to some extent.  The normalization of the Ginzburg number in the
theoretical expressions has been chosen such that the curves coincide in the
asymptotic (Ising) regime. The main remaining difference then essentially lies
in the width of the crossover regime, where the asymptotic crossover described
by the field-theoretic curve pertains to the strict limit $G \to 0$. Systems
with a finite Ginzburg number will appear to exhibit a {\em faster\/}
crossover, due to the fact that one leaves the critical region before
completing the full crossover and hence enters the noncritical high-temperature
regime, as has e.g.\ been illustrated in Fig.~4 of Ref.~\cite{chicross}.
Additional deviations, however, may result from neglected irrelevant couplings
and analytic corrections to scaling~\cite{bagnuls87} and it remains to be seen
which theoretical expression actually provides the best description of
experimental measurements.  The admirable accuracy of the Xe data from Refs.\
\cite{Guettinger:C:81} and~\cite{Michels:W:L:54}, and the large temperature
range that is spanned by the total data set, enable an accurate matching with
the theoretical expressions. As can be seen from Fig.~\ref{fig:compr}, the LB
and BK calculations turn out to yield a strikingly good description of the data
over the full temperature range. The BB expression exhibits a slower crossover
toward the classical critical regime, as was already found in
Ref.~\cite{bagnuls84b}. We note here that the agreement is very sensitive to
the choice of the critical amplitude; the value $\Gamma^+ = 0.0577 \pm
0.0001$~\cite{Guettinger:C:81} for $\gamma=1.241$ appears very low. Upon
omission of the data points for $t > 0.02$, which are presumably not described
by the first few Wegner corrections, a fit yields $\Gamma^+ = 0.0587 \pm
0.0002$; for $\gamma = 1.240$ this even increases to $\Gamma^+ = 0.0594 \pm
0.0002$.  A smaller value improves the agreement with the BB curve for
intermediate temperatures, but leads to poor agreement for very small~$t$.  It
should be stressed that the horizontal offset between the curves and the
experimental data should {\em not\/} be regarded as a failure by itself, since
the matching involves adjustment of the Ginzburg parameter~$G$, here estimated
as $G^{+}_{\rm Xe} = 0.018 \pm 0.002$.  The essential point is that there
exists no value of $G$ for which the experimental data coincide with the BB
curve over the full temperature range. Note that even the data point at the
highest temperature is located at $t < 0.5$, and thus lies definitely not far
outside the critical regime.

The data for $^3$He (also shown in Fig.~\ref{fig:compr}) exhibit a completely
different behavior. Whereas the initial increase of the data is again well
described by the BK and LB calculations, with $G^{+}_{\rm He} = 0.0025 \pm
0.0010$, a systematic deviation sets in around $t = 0.005$, which is not
captured by any of the theoretical expressions. The experimental data continue
their upward trend, but at a markedly slower rate. Unlike for $t<0$, there is
no reason to expect a deviation at such a small~$t$.  In the rightmost part of
the graph the data run roughly parallel to the predicted curves, but shifted by
an order of magnitude in $t$. At this point, the system has obviously left the
critical region and one is observing the changeover to regular high-temperature
behavior. This effect by itself, however, would rather {\em hasten\/} the
(apparent) crossover and can certainly not explain the slowing down in the
crossover behavior; the Xe results (with a 7 times higher $G$!) only reinforce
this conclusion.

We suggest that one is actually observing the interplay between the critical
and the quantum fluctuations. While the former are dominating close to~$T_c$
(and thus lead to Ising-like critical behavior), the contribution of the latter
cannot be neglected in this temperature range and leads to a considerable
enhancement of the compressibility, as follows from the critical amplitude
$\Gamma^+ = 0.139 \pm 0.003$, which is more than twice as large as for Xe.
Indeed, the thermal de Broglie wavelength~$\lambda_{\rm dB}$ is, at $T_c$, 4.5
times larger than the van der Waals radius, compared to a corresponding ratio
of only 0.062 for Xe at criticality.  However, due to the low value of~$T_c$,
the absolute temperature changes appreciably within the crossover region and so
does the quantum-natured contribution to the compressibility. Since the data in
Fig.~\ref{fig:compr} are normalized by the critical amplitude, this then
obviously leads to an apparent {\em depression\/} of the compressibility at
higher $t$ and to a corresponding ``widening'' of the crossover regime.
Although a proper description of crossover behavior is clearly beyond the
scope of the van der Waals equation, it is still instructive to consider the
quantum corrections to the second virial coefficient~\cite{Hirschfelder}. In
first approximation, these corrections are indeed found to be proportional to
$\lambda_{\rm dB}^2 \propto T^{-1}$.  Exchange effects, which will tend to {\em
decrease\/} the compressibility for $^3$He, are proportional to $\lambda_{\rm
dB}^3$ but expected to become important only below 1~K.  The compressibility of
$^4$He~\cite{helium4} lends strong support to this scenario, as it shows
essentially the same behavior as $^3$He, except that the high-temperature data
lie somewhat higher, i.e., closer to the theoretical curve. This is consistent
with the reduced importance of quantum effects in $^4$He due to its higher
mass. A quantitative description of this phenomenon will require a more
fundamental understanding of the contribution of quantum-mechanical
fluctuations to the critical divergence of the compressibility, which is
essentially driven by {\em thermal\/} fluctuations!  Of particular interest
will be a study of the compressibility in the two-phase region, where the
above-mentioned mechanism should rather have an opposite effect.

For completeness we remark that a rather different fit of the $^3$He data to
the Monte Carlo results might be attempted, in which a considerably higher
critical amplitude $\Gamma^+ \approx 0.15$ is used. This leads to a larger
Ginzburg number and diminishes the deviations at higher~$t$, but entails two
undesirable consequences: $\Gamma^+$ lies certainly three standard deviations
above the value quoted in Ref.~\cite{Pittman:D:M:79} and a considerable part of
the data at very small~$t$ has to be omitted from the analysis, even though
there is no reason to expect stratification effects at these temperatures.
With regard to the Ginzburg numbers found in our analysis, we note that the
$G^{+}_{\rm Xe}/G^{+}_{\rm He} > 1$, whereas $G^{-}_{\rm Xe}/G^{-}_{\rm He}$
(as obtained from the coexistence curve) lies much closer to unity; this is
consistent with the corresponding first Wegner corrections, to which the
Ginzburg parameter can be related, but seems to contradict the expected
universality in the ratio of Ginzburg numbers above and
below~$T_c$~\cite{aharony80}.

%\acknowledgments
We wish to thank H. Carr and M. Durieux for supplying us with
data tabulations for Xe and $^4$He, A. Kogan for help with scaled plots, and
K. Binder, M. Anisimov and M. E. Fisher for stimulating comments and helpful
discussions. The provision of computing resources by the NIC and support by
NASA grant NAG3-1838 are gratefully acknowledged.

%\bibliographystyle{prsty}
%\bibliography{crossover}

\end{document}